\documentclass[notitlepage,nofootinbib,showpacs,unsortedaddress,aps,prd, 12pt]{revtex4-1}
\usepackage[utf8x]{inputenc}
\usepackage{amsfonts,amsmath, bm,bbold,latexsym,amssymb,slashed, graphicx}
\usepackage[normalem]{ulem}
\usepackage{color}
\usepackage{hyperref}

\numberwithin{equation}{section}

\newcommand{\ins}[1]{}

\newcommand{\Tr}{\ensuremath{\,\text{Tr}\,}}

\newcommand{\vev}[1]{\ensuremath{\left< #1 \right> }}

\newcommand{\half}{{\textstyle{\frac 1 2}}}

\newcommand{\vac}{\ensuremath{|\text{vac}\rangle}}
\newcommand{\vevv}[1]{{\ensuremath{\langle\text{vac}|}#1{|\text{vac}\rangle}}}

\newcommand{\eps}{\varepsilon}

\newcommand{\beq}{\be}
\newcommand{\eeq}{\ee}
\newcommand{\p}{\partial}
\newcommand{\beqa}{\begin{eqnarray}}
\newcommand{\eeqa}{\end{eqnarray}}

\newcommand{\tcR}{\widetilde{\cal R}}

\newcommand{\ket}[1]{\ensuremath{\vert #1 \rangle}}

\newcommand{\drop}[1]{}
\newcommand{\be}{\begin{equation}}
\newcommand{\ee}{\end{equation}}
\renewcommand{\eqref}[1]{Eq.~(\ref{#1})}

\renewcommand{\L}{\ensuremath{\mathcal L}}

\begin{document}

\title{Poincar{\'e} Symmetry of the GZ-Model}

\author{Martin Schaden}
\email{mschaden@rutgers.edu}
\affiliation{Department of Physics, Rutgers, The State University of New Jersey, 101 Warren Street, Newark, New Jersey - 07102, USA }

\author{Daniel Zwanziger}
\email{dz2@nyu.edu}
\affiliation{Physics Department, New York University, 4 Washington Place, New York, NY 10003, USA}

\begin{abstract}
\noindent{\bf Abstract:} Due to internal symmetries of its ghost sector, the Poincar{\'e} generators of the Gribov-Zwanziger (GZ) model are not unique.  The model apparently has two linearly independent  symmetric and conserved energy momentum tensors.  We show that these energy-momentum tensors are physically equivalent and differ by unobservable conserved currents only. There is a single physical energy-momentum operator that is invariant under all symmetries of the ghost sector, including Becchi-Rouet-Stora-Tyutin (BRST).  This resolves concerns about Poincar\'e invariance raised by the explicit $x$-dependence of the BRST operator.  The energy, momentum and angular momentum of physical states are well-defined quantities that vanish for the ground state of this theory. We obtain and discuss the physical Ward identities resulting from Poincar{\'e} invariance.     
\end{abstract}
\maketitle

\section{Introduction} 
The GZ-theory is a non-perturbative approach to QCD that self-consistently accounts for infrared effects. The theory is defined by a local and renormalizable continuum action that depends on a dimensionful parameter $\gamma$, the Gribov mass.  This model was originally proposed~\cite{Zwanziger:1989mf} to dynamically constrain a gauge  theory to the first Gribov region~\cite{Gribov:1977wm} and the gap equation that determines the value of the Gribov parameter $\gamma$, has become known as the ``horizon condition".  In~\cite{I} the value of $\gamma$ was related to the trace anomaly of the energy momentum tensor. 

The nilpotent BRST symmetry of the GZ-theory is manifestly broken by, for instance, the non-vanishing expectation value ,
\beq
\label{swbar}
\langle \ s \partial_\mu\bar\omega^a_{\nu b}(x) \ \rangle=\gamma^{1/2} \delta_{\mu\nu}\delta^a_b\neq 0, 
\eeq   
where $\bar\omega^a_{\mu b}$ is one of the auxiliary ghost fields of the theory. If we assume the existence of a well-defined BRST charge $Q_B$ that effects the $s$-operation, $\{ Q_B, \bar\omega^a_{\mu b}\} = s \bar\omega^a_{\mu b}$,  and a vacuum state $\vac$, the expectation value  $\vevv{\{ Q_B, \partial_\mu\bar\omega^a_{\nu b}(x) \}} \neq 0$ formally implies that $Q_B \vac\neq 0$. 

Compared to perturbative Faddeev-Popov theory, the GZ-model not only includes a  large number of additional unphysical fields but also a large number of  symmetries under which physical observables are required to be invariant.  Since these symmetries (including BRST)  and their breaking are not observable as multiplets of degenerate physical excitations,  we called them phantom symmetries in~\cite{I}. For convenient reference, the charges of some of these phantom symmetries are given in Appendix~\ref{phantomsymm}. We refer the reader to the 
appendix of~\cite{I} for the definitions of the full set of phantom symmetry generators, 
\beq\label{phantomset}
\mathfrak{F}=\{Q_B,Q_\mathcal{N},Q_+, Q_{\bar\varphi,\mu},Q_{\bar\omega,\mu},Q^a_G,Q^a_C,Q^a_{NL},Q^a_{\bar c},Q_{R,\mu a},Q_{S,\mu a},Q_{E,\mu\nu\,ab},Q_{F,\mu\nu\,ab},Q^a_{U,\mu b},Q^a_{V,\mu b}\}\ ,
\eeq
of this model.  For a review of earlier work, see~\cite{Vandersickel:2012tz}.

In~\cite{I} we reconstructed the physical states of the theory from the physical observables $F$, 
\beq\label{W}
\mathcal{W}_{\rm phys}=\{F; [Q_X,F]=0, \text{for all } Q_X\in\mathfrak{F}\}\ ,
\eeq
that commute with all phantom symmetry generators $Q_X$. The expectation value of a functional $F\in\mathcal{W}_{\rm phys}$ by this definition does not depend on the choice of vacuum state. With the hypothesis of~\cite{I} that BRST-exact physical observables  have vanishing  expectation value, 
\beq\label{hypothesis}
sF\in\mathcal{W}_{\rm phys}\Rightarrow \vev{sF}=0\ ,
\eeq  
BRST-exact physical states form a null-space, and GZ-theory provides a quantization of classical Yang-Mills theory. If this hypothesis holds, the spontaneous breaking of phantom symmetries is segregated from the physical sector.  

However, the phantom symmetries allow the construction of a second linearly independent conserved symmetric tensor.  It is composed of conserved Noether currents that correspond to a Poincar\'e algebra of internal phantom symmetries.   Since any linear combination of the two conserved symmetric tensors potentially is an EM-tensor, the definition of energy, momentum and angular momentum in this theory appears ambiguous. We shall see that this is not the case. Energy, momentum and angular momentum in fact operate uniquely on $\mathcal{W}_{\rm phys}$. Perhaps surprisingly, these generators  are not the canonical ones of translations and rotations.

When the horizon condition is fulfilled, all vacua of the model have the same energy and are degenerate. They break BRST and other phantom symmetries, but are physically equivalent and have vanishing momentum and angular momentum. In view of \eqref{swbar} this claim may appear preposterous. It holds because the canonical generators of unbroken translations,  $\widehat{P}_\mu$, do not commute with the BRST-charge $Q_B$ of this model. Despite its spontaneously broken BRST-symmetry, the GZ-theory thus is remarkably consistent. 

This article is organized as follows. In the next section we introduce the GZ-model and fix notation. We discuss the horizon condition and  BRST-symmetry of the theory.  Sect.~\ref{Poincaresym} examines the generators of translations and Euclidean rotations of the model and their relation to other generators and in particular BRST:  in Sect.~\ref{canonicalGen} we find that the canonical generators of space-time isometries do not commute with the BRST-charge $Q_B$ and other phantom charges of the model; the Poincar\'e algebra that commutes with all generators of unphysical symmetries is found in Sect.~\ref{invGen} and related to the canonical one; Sect.~\ref{physPoincare} examines how space-time isometries transform physical states of the theory. The (conserved) energy-momentum tensor is discussed in Sect.~\ref{emGZ}. We consider the physical energy momentum tensor of~\cite{I} in Sect.~\ref{physT} and relate it to the energy momentum tensor of Belinfante in Sect.~\ref{BelinfanteT}. The difference between the two tensors is a symmetric tensor that is conserved up to equations of motion of unphysical fields only. Its expectation value vanishes due to the horizon condition.  The energy-momentum and angular momentum Ward identities are derived in Sect.\ \ref{WardId}, and we show that they reduce to those of Faddeev-Popov theory for physical operators provided that the hypothesis of \eqref{hypothesis} holds. In Sect.~\ref{alternateview} we derive an alternate, but equivalent, form of the Ward identity by a different method and show that it too reduces to the physical one when the hypothesis holds.  Sect.~\ref{diss} summarizes and discusses our results. The  phantom symmetry generators we use are compiled in App.~\ref{phantomsymm} and the energy-momentum tensor of Belinfante of the GZ-theory is derived in App.~\ref{Belinfante}.     

\section{The local GZ-action and spontaneously broken BRST}
\noindent The GZ-model proposed in~\cite{Zwanziger:1989mf} is defined by the Euclidean continuum action,
\beq
\label{SGZ}
S\equiv \int d^dx {\cal L} \equiv \int d^dx ({\cal L}^{\rm YM} + {\cal L}^{\rm gf})\ ,
\eeq
where ${\cal L}^{\rm YM}= \frac{1}{4} (F_{\mu \nu})^2$, with $F_{\mu \nu}= \p_\mu A_\nu - \p_\nu A_\mu + A_\mu\times A_\nu$, is the  gauge-invariant Yang-Mills Lagrangian with gauge potential $A_\mu$. Here and in the following we suppress the gauge coupling constant and explicit adjoint color indices when possible. Associating upper Latin indices with color and lower Latin indices with `flavor', we adopt  the following notation~\cite{I,Mader:2014}: $(X\times Y)^a\equiv g f^{abc} X^b Y^c$ and $(X\widetilde{\times} Y)^a\equiv g f^{abc} X_b Y_c$. $\text{Tr} X\equiv\sum_a X^a_a$ is the diagonal trace of the matrix $X$.  We also introduce a dot product $X\cdot Y=\sum_{ab}X_b^a Y_b^a$ that sums over color and, if applicable, flavor. The gauge covariant derivative of the adjoint representation in this notation is $D_\mu X \equiv \p_\mu X + A_\mu\times X$. 

The local gauge-fixing part ${\cal L}^{\rm gf}$ of the GZ-model is given by,
\beqa
{\cal L}^{\rm gf} & = & i \p_\mu b\cdot A_\mu  - i \partial_\mu \bar c \cdot D_\mu c
+ \p_\mu \bar\varphi_{\nu b}\cdot D_\mu \varphi_{\nu b} 
- \p_\mu \bar\omega_{\nu b}\cdot [D_\mu \omega_{\nu b}  + (D_\mu c )\times \varphi_{\nu b}]
\nonumber \\
\label{Lgf}
 &&+ \gamma^{1/2}\text{Tr} [ D_\mu ( \varphi_\mu - \bar\varphi_\mu) -  (D_\mu c) \times \bar\omega_\mu] - \gamma d (N^2 -1)\ . 
\eeqa
The first two terms of ${\cal L}^{\rm gf}$ are familiar from perturbative Faddeev-Popov theory in Landau gauge. The additional terms involve auxiliary fermionic and bosonic ghost fields $\varphi, \omega,\bar\omega,\bar\varphi$. The nilpotent BRST-variation of the fields is given by, 
\begin{align}
\label{sonfields}
s A_\mu & = D_\mu c  & s c & = - \half c \times c
\nonumber   \\
s {\bar c} & =   b   &  s b &= 0
\nonumber \\
s \varphi^a_{\mu b} & = \omega^a_{\mu b}  & s \omega^a_{\mu b} &= 0 \ 
\nonumber \\
\ s \bar\omega^a_{\mu b} & =   \bar\varphi^a_{\mu b} + \gamma^{1/2} x_\mu \delta^a_b  & s \bar\varphi^a_{\mu b} &= 0,
\end{align}
and ${\cal L}^{\rm gf}$ formally is a BRST-exact extension of the Yang-Mills Lagrangian given by the gauge-fixing fermion $\Psi$,
\beq
\label{Psi}
{\cal L}^{\rm gf}=s\Psi,  \ \ \text{with}\ \Psi \equiv  i \p_\mu {\bar c} \cdot A_\mu + \p_\mu \bar\omega_{\nu}\cdot D_\mu \varphi_{\nu} - \gamma^{1/2} \text{Tr} D_\mu\bar\omega_\mu\ .
\eeq
The generator of BRST transformations corresponding to \eqref{sonfields} is,
\beq
\label{QB}
Q_B=\int d^dx \ \text{Tr}\left[ (D_\mu c)\cdot {\delta \over \delta A_\mu} - \half (c \times c) \cdot {\delta \over \delta c} + b\cdot  {\delta \over \delta {\bar c}} + \omega_\mu\cdot {\delta \over \delta \varphi_\mu} + \bar\varphi_\mu\cdot  {\delta \over \delta {\bar\omega}_\mu}+ \gamma^{1/2} x_\mu\text{Tr}  {\delta \over \delta {\bar\omega}_\mu} \right].
\eeq
Note that the inhomogeneous term proportional to $x_\mu$ in the BRST-variation of $\bar\omega^a_{\mu b}$ in \eqref{sonfields} does not lead to an explicit coordinate dependence of ${\cal L}^{\rm gf}$ in \eqref{Lgf},  because the gauge fermion $\Psi$ depends on $\partial_\mu \bar\omega^a_{\nu b}$ and $\text{Tr} A_\mu\times \bar\omega_\mu$ only.

For $\gamma\neq 0$, the inhomogenous transformation of $\bar\omega$ in \eqref{sonfields} implies that this BRST-symmetry of the theory is spontaneously broken.\footnote{Another approach considers the same theory from the point of view of a different, explicitly broken, BRST symmetry~\cite{Sorella:2004}. The soft breaking of BRST has been extended to include dimension two condensates~\cite{Dudal:2005}.} Requiring the effective action to be stationary with respect to $\gamma$ at vanishing quantum fields gives the "horizon condition", 
 \beq
\label{horizon}
\left\langle \text{Tr}[ D_\mu (\varphi_\mu - \bar\varphi_\mu) - \bar\omega_\mu\times D_\mu c  ] \right\rangle = 2\gamma^{1/2} d (N^2 -1) ,
\eeq 
which is a gap equation that implicitly determines~\cite{Zwanziger:1989mf, Schaden:1994, Gracey:0510} the Gribov mass $\gamma$. \eqref{horizon} was originally~\cite{Zwanziger:1989mf} derived as a necessary condition for field configurations in Landau gauge to be dynamically constrained to the first Gribov region. The calculation of the trace anomaly in~\cite{I} confirms that a non-vanishing positive value of $\gamma$ is energetically favored and that BRST in this model thus is spontaneously broken.\footnote{To derive local Ward identities, it is sufficient~\cite{Vandersickel:2012tz} that the local generator $\delta_\epsilon \equiv \epsilon(x) s$ exists and acts on the Lagrangian density according to
$\delta_\epsilon {\cal L} = j_{{\rm BRST},\mu} \ \p_\mu \epsilon$.}

[The theory with $\gamma \neq 0$ can be formally obtained from the theory with $\gamma = 0$ by  a shift of the bose ghost fields $\varphi_{\mu b}^a(x) \to \varphi_{\mu b}^a(x) - \gamma^{1/2} x_\mu \delta_b^a$ and $\bar\varphi_{\mu b}^a(x) \to \bar\varphi_{\mu b}^a(x) + \gamma^{1/2} x_\mu \delta_b^a$  \cite{Schaden:1994}.  (The explicit $x$-dependence cancels out of the action.)  In this way the shifted action inherits all the obvious symmetries of the unshifted action, which are otherwise not so easily found by inspection of the shifted action.  (These symmetries are given in the appendix of~\cite{I}.)  However, once found, they are true and valid symmetries of the (shifted) action,  \eqref{Lgf}, as one may verify by direct calculation, without introducing the unshifted action or the shift.  The symmetries are transformations of fields within a Hilbert space, whereas the shift takes one to an orthogonal space.  For this reason, in the present article, the shift and the unshifted fields make no appearance (outside this paragraph).]

On a finite volume with periodic boundary conditions for all fields, the transformation of \eqref{sonfields} has to be modified since $x_\mu$ is not a periodic function. This clarifies the spontaneous breaking of the fermionic BRST-symmetry of this theory and gives a criterion for determining whether the expectation value of a local BRST-exact functional vanishes in the thermodynamic limit. It  was proven~\cite{I} that $\vev{sF}=0$ for a physical observable $F$ with local support, if in the continuum limit 
\beq\label{criterion}
\lim_{|x|\rightarrow\infty} |x|^d \vev{F\   s \text{Tr} D_\mu\varphi_\mu(x)}=\lim_{|x|\rightarrow\infty} |x|^d \vev{F\   \text{Tr} (D_\mu\omega_\mu+D_\mu c\times \varphi_\mu)(x)}=0\ ,
\eeq
where $d$ is the dimension of Euclidean space-time. This in particular generally holds if the correlation function of \eqref{criterion} is regular in momentum space at $p^2=0$. [The criterion of \eqref{criterion} in this sense expresses the requirement that the state $F\vac$ in canonical quantization \emph{does not} couple to massless (Goldstone) excitations of the spontaneously broken BRST symmetry].  

Remarkably, the criterion of \eqref{criterion} for instance can be used to show~\cite{I} that,
\beq\label{halfhorizon}
0=\langle s\,\text{Tr} D_\mu\bar\omega_\nu\rangle = \langle\text{Tr} [ D_\mu \bar\varphi_\nu+ \bar\omega_\nu\times D_\mu c ]  \rangle +\gamma^{1/2}(N^2-1) \delta_{\mu\nu},
\eeq
when \eqref{horizon} is satisfied. \eqref{halfhorizon} will establish that the energy density of the vacuum is well defined. Together with Euclidean rotation symmetry, \eqref{halfhorizon} implies the stronger statement,
\beq\label{halfhorizon2}
\langle\text{Tr} [ D_\mu \varphi_\nu]  \rangle =-\langle\text{Tr} [ D_\mu \bar\varphi_\nu+ \bar\omega_\nu\times D_\mu c ]  \rangle = \gamma^{1/2}(N^2-1) \delta_{\mu\nu},
\eeq
when the horizon condition of \eqref{horizon} is satisfied.

\section{Poincar{\'e} symmetry}
\label{Poincaresym}
Although the theory has been continued to Euclidean space-time, we ignore the signature of the metric in most of the following and collectively refer to translations and SO(4)-rotations of Euclidean space-time as Poincar\'e symmetries and the algebra of their generators as a Poincar\'e algebra.
 
\subsection{Canonical Poincar{\'e} Generators}
\label{canonicalGen}
The canonical generators of translations, $\widehat{\cal P}_\nu$, of this model are, 
\beq\label{hatP}
\widehat{\cal P}_\nu = \int d^dx \ \widehat{\mathfrak{p}}_\nu,
\eeq
with the densities, 
\beqa
\label{phat}
\widehat{\mathfrak{p}}_\nu \equiv \p_\nu A_\mu\cdot {\delta  \over \delta A_\mu} + \p_\nu b\cdot  {\delta \over \delta  b}  + \p_\nu c\cdot {\delta \over \delta c} 
  + \p_\nu {\bar c}\cdot {\delta \over \delta {\bar c}} + \p_\nu \varphi_{\mu }\cdot  {\delta \over \delta \varphi_{\mu}} 
 \nonumber \\
+ \p_\nu \bar\varphi_{\mu } \cdot{\delta \over \delta \bar\varphi_{\mu }}  + \p_\nu \omega_{\mu } \cdot{\delta \over \delta \omega_{\mu }}  + \p_\nu {\bar\omega}_{\mu}\cdot  {\delta \over \delta {\bar\omega}_{\mu}}.
\eeqa
Due to the interaction $\gamma^{1/2}\text{Tr} D_\mu (\varphi_\mu - \bar\varphi_\mu)$ in $\mathcal{L}^{\rm gf}$ of \eqref{Lgf}, the auxiliary ghost fields canonically transform as vectors under Euclidean rotations. The corresponding canonical angular momentum generators are,
\beq
\label{Mhat}
\widehat{\cal M}_{\mu \nu} = \int d^dx \ \left[x_\mu \widehat{\mathfrak{p}}_\nu  + A_\nu \cdot {\delta \over \delta A_\mu}  + \mathfrak{s}_{\mu \nu}\right]-[\mu\leftrightarrow\nu],
\eeq
with 
\beq
\label{sdensity}
\mathfrak{s}_{\mu \nu} = \varphi_{\nu}\cdot {\delta  \over \delta \varphi_{\mu}}
+ \bar\varphi_{\nu}\cdot  {\delta  \over \delta \bar\varphi_{\mu}} 
 + \omega_{\nu}\cdot  {\delta  \over \delta \omega_{\mu}} 
      + {\bar\omega}_{\nu}\cdot  {\delta  \over \delta {\bar\omega}_{\mu}} \ .
\eeq
The operators $\widehat{\cal P}_\mu$ and $\widehat{\cal M}_{\mu\nu}$  are symmetries of the action,
\beq
\label{commutewS}
[ \widehat{\cal P}_\mu, S] = [ \widehat{\cal M}_{\lambda \mu}, S] = 0.
\eeq
and satisfy the commutation relations of a (Euclidean) Poincar\'e algebra,
\beqa
\label{hatcommutePoincare}
[ \widehat{\cal P}_\mu, \widehat{\cal P}_\nu ] &=& 0;  \ \ \ \ \ \  [ \widehat{\cal M}_{\lambda \mu}, \widehat{\cal P}_\nu ] = \delta_{\lambda \nu} \widehat{\cal P}_\mu - \delta_{\mu \nu} \widehat{\cal P}_\lambda\nonumber\\
 ~[ \widehat{\cal M}_{\lambda \mu}, \widehat{\cal M}_{\sigma \tau} ] &=& \delta_{\lambda \sigma} \widehat{\cal M}_{\mu \tau} - \delta_{\mu \sigma} \widehat{\cal M}_{\lambda \tau} - \delta_{\lambda \tau} \widehat{\cal M}_{\mu \sigma} + \delta_{\mu \tau} \widehat{\cal M}_{\lambda \sigma} .
\eeqa
Remarkably, the BRST-operator $Q_B$ of \eqref{QB} does not commute with the canonical momentum operators $\widehat{\cal P}_\mu$ of this theory. We find that,
\beq
\label{commPQ}
[Q_B,\widehat{\cal P}_\mu]=\gamma^{1/2}\int d^dx \ \text{Tr} {\delta \over \delta {\bar\omega}_\mu}=\gamma^{1/2} Q_{\bar\omega,\mu} \ ,
\eeq  
where $Q_{\bar\omega,\mu}\in \mathfrak{F}$ is the charge of a phantom symmetry ( see \eqref{Qw} of Appendix~\ref{phantomsymm}). 

Jacobi's identity implies that \eqref{swbar} can be written in the form,
\beqa
\label{choice}
\langle \ s \partial_\mu\bar\omega^a_{\nu b}(x) \ \rangle = \langle \ \{Q_B,[\widehat{\cal P}_\mu,\bar\omega^a_{\nu b}(x)]\}\ \rangle &=&\langle \ [\widehat{\cal P}_\mu,\{Q_B, \bar\omega^a_{\nu b}(x)\}]\ \rangle+\langle \ \{[Q_B,\widehat{\cal P}_\mu],\bar\omega^a_{\nu b}(x)\}\ \rangle\nonumber\\
&=&\langle \ [\widehat{\cal P}_\mu,\{Q_B, \bar\omega^a_{\nu b}(x)\}]\ \rangle+\gamma^{1/2}\delta_{\mu\nu}\delta^a_b\ , 
\eeqa   
where \eqref{commPQ} was used to evaluate the last contribution.  The fact that for  $\gamma> 0$, $\widehat{\cal P}_\mu$ and $Q_B$  commute to a broken phantom symmetry already implies that the theory cannot be invariant under \emph{both}, translations \emph{and} BRST.  

Unbroken invariance of an Euclidean field theory under a continuous symmetry generated by a hermitian bosonic charge $Q$, is the statement that for any real parameter $\lambda$ and local functional $F$ of the fields,
\begin{align}
\label{symmQ}
\vev{F}&\equiv\vevv{F}=\vevv{e^{i\lambda Q} F e^{-i\lambda Q}}\nonumber\\
&=\vev{F}+i\lambda \vev{[Q,G(\lambda)]}\nonumber\\
&\text{with }G(\lambda)=F+i\frac{\lambda}{2!}[Q,F]-\frac{\lambda^2}{3!}[Q,[Q,F]] +\dots\ .
\end{align}
This notion of an unbroken symmetry is readily extended to fermionic charges $Q$ and a graded algebra by introducing anti-commuting Grassmann parameters $\lambda$. With this extension \eqref{symmQ} implies that,
\beq\label{Qunbroken}
Q\ \text{\emph{generates an unbroken continuous symmetry}}\Leftrightarrow \vev{[\lambda Q,F]}=0\ ,
\eeq
for \emph{all} local functionals $F$ of the Euclidean field theory. A symmetry thus is unbroken if the expectation value of the  graded commutator of every local functional with its generator vanishes.  Since this holds for every local functional $F$, \eqref{symmQ} and \eqref{Qunbroken} may formally be viewed as requiring that,
\beq\label{formalinv}
\ket{\lambda}=e^{-i\lambda Q}\vac=\vac \ \text{or that }Q\vac=0\ ,
\eeq
for an unbroken symmetry. In the Euclidean theory \eqref{formalinv} is somewhat formal.  According to the reconstruction theorem\cite{Wightman:1956,I},  the action of a functional derivative operator (like the charge) on a state of the theory represented by the functional $F$, is given by $Q\ket{F}\equiv[Q,F]\vac=\ket{[Q,F]}$. The vacuum is represented by the unit functional and thus $Q\vac\equiv[Q,1]\vac=0$ necessarily vanishes.  However, the reconstruction theorem \emph{presumes} that the ground state is unique. In a theory with a spontaneously broken symmetry, the representation of the vacuum state by unity makes sense only for the restricted set of (invariant) physical observables~\cite{I}. \eqref{symmQ} and \eqref{Qunbroken} do not require this restriction and allow us to discuss spontaneous symmetry breaking on the full space of \emph{all} functionals of the theory.    

We require the theory to be translationally invariant. The ground state, $\vac$, is translationally and rotationally invariant if,
\beq\label{symmvac}
\widehat{P}_\mu\vac=\widehat{M}_{\mu\nu}\vac =0\ ,
\eeq
in the sense (just discussed) that the expectation value of the commutator of any local functional with the canonical momentum or angular momentum vanishes. 
For $\gamma>0$,  \eqref{choice} then implies that this vacuum is BRST variant, 
\beq\label{brokenBRST}
Q_B\vac\neq 0\ ,
\eeq
because the expectation value of the commutator of $Q_B$ with $[\widehat{\cal P}_\mu,\bar\omega^a_{\nu b}(x)]=\partial_\mu\bar\omega^a_{\nu b}(x)$ does not vanish.

In the sense of \eqref{symmQ}, $|\lambda\rangle\equiv e^{-i\lambda Q_B}\vac =\vac +i\lambda Q_B\vac$ are (formal) states for which  $\langle\lambda |F|\lambda\rangle=\langle{\rm vac}|F\vac$ for any physical observable $F\in\mathcal{W}_{\rm phys}$. We have that the expectation value of any BRST-invariant functional is the same in all these states,
\beq\label{vevexactdep}
\langle\lambda | F|\lambda\rangle=\vevv{ F}+i\vevv{ [\lambda Q_B,F]}=\vevv{F} .
\eeq
This in particular holds for any BRST-exact functional $F=sX$ and implies that BRST is spontaneously broken in any state $\ket{\lambda}$ if  it is broken in one of them. The horizon condition of \eqref{horizon} also is satisfied for all states $\ket{\lambda}$ if it holds in one of them.  

However, the formal states $\ket{\lambda}$ do not all have vanishing ghost number $Q_{\cal N}$ and momentum $\widehat{P}_\mu$. We have that  $Q_{\cal N}|\lambda\rangle=-i\lambda Q_B\vac\neq 0$  and $\widehat{P}_\mu|\lambda\rangle=i\lambda \gamma^{1/2} Q_{\bar\omega,\mu}\vac\neq 0$  if $\vac$ has vanishing ghost number and satisfies \eqref{symmvac}. In particular,
\beqa\label{mom}
\langle\lambda |\ [\widehat P_\mu,\bar\omega_\nu]\ |\lambda\rangle &=&\langle{\rm vac}|\ [\widehat P_\mu,\bar\omega_\nu]\ \vac+i\lambda\langle{\rm vac}|\ \{Q_B,[\widehat{\cal P}_\mu,\bar\omega^a_{\nu b}(x)]\}\ \vac\nonumber\\
&=&\langle{\rm vac}|\ [\widehat{\cal P}_\mu,\bar\omega_\nu+i\lambda \{Q_B, \bar\omega^a_{\nu b}(x)\}]\ \vac+i \lambda \gamma^{1/2}\delta_{\mu\nu}\delta^a_b\ , 
\eeqa
where we again used Jacobi's identity as in \eqref{choice}.    

Requiring that ghost number and translational symmetry remain unbroken thus constrains the number of admissible vacua. Even though the parameter $\lambda$ here is anti-commuting, this is somewhat analogous to the spontaneous breaking of a global bosonic symmetry like $O(n)$: $O(n)$-symmetric functionals have the same expectation value for any degenerate vacuum, but only some of the vacua are invariant under subgroups of $O(n)$ that are not broken.  

In the following we assume that the GZ-theory has a (unique) ground state $\vac$ that in addition to \eqref{symmvac} also satisfies,
\beq\label{invvac}
Q_{\cal N}\vac=Q_+\vac=Q_{R,\mu a}\vac=Q^a_C\vac=0\ ,
\eeq
in the sense that the vacuum expectation value of the (graded) commutator of any functional with these charges vanishes. All other phantom symmetries  apparently are spontaneously broken for $\gamma>0$.  One of the more interesting unbroken phantom symmetries is generated by $Q_{R,\mu a}$, defined in \eqref{QR}. This symmetry arises due to the absence of a  vertex containing $\bar c$ and $\omega$ in ${\cal L}^{\rm gf}$ and implies that any functional with a positive number of FP-ghosts, $\#c>\#\bar c$, has vanishing expectation value. The $c-\bar\omega$-propagator of this theory thus vanishes. An unbroken $Q_R$-symmetry implies that the propagator of the auxiliary fermi ghosts and of the FP-ghosts are related~\cite{I},
\beq\label{auxFP}
0=\vev{[Q_{R,\mu a}, \omega^b_{\nu c}(x) \bar c^d(y)]}_A=\delta_{ac}\delta_{\mu\nu}\vev{  c^b(x) \bar c^d(y)}_A+i\vev{\omega^b_{\nu c}(x) \bar\omega^d_{\mu a}(y)}_A\ ,
\eeq  
where $\vev{\dots}_A$ denotes the expectation value for a fixed background connection. 

\subsection{Invariant Poincar\'e Generators}
\label{invGen}
Since the generators of translations $\widehat{\cal P}_\mu $ do not commute with $Q_B$  and other phantom symmetries, it is not immediatly apparent that $\mathcal{W}_{\rm phys}$ is closed under space-time isometries. We here show that it is. We also show this for rotations.

In~\cite{I} we considered a Poincar\'e algebra of operators that commute with the BRST charge $Q_B$ of \eqref{QB}. The momenta ${\cal P}_\mu$ of this invariant Poincar\'e algebra are,
\beq
\label{P}
{\cal P}_\mu = \widehat{\cal P}_\mu+\gamma^{1/2} (Q_{\bar\varphi,\mu}-Q_{\varphi,\mu})\ .
\eeq
For $\gamma>0$, the ${\cal P}_\mu$ do not simply generate space-time translations of the ghost fields. However, the additional conserved charges are associated with internal color-singlet phantom symmetries. They are given in \eqref{Qphibar} and \eqref{Qphi} of Appendix~\ref{phantomsymm}.

The angular momenta of this invariant Poincar\'e algebra similarly differ from the canonical generators of Euclidean rotations of \eqref{Mhat} by phantom charges only,
\beq\label{M}
{\cal M}_{\mu\nu} = \widehat{\cal M}_{\mu\nu} -Q_{M,\mu\nu} .
\eeq
The generators of internal $SO(4)$-rotations of auxiliary ghosts, $Q_{M,\mu\nu}$, are given in \eqref{QM} of Appendix~\ref{phantomsymm}.  Note that the auxiliary fields of the model do not transform as vectors and that ${\cal M}_{\mu\nu}$ in this sense does not simply generate rotations of Euclidean space-time even for $\gamma\rightarrow 0$.

The invariant generators satisfy a Poincar\'e algebra of their own,
\begin{align}
\label{commutePoincare}
[ {\cal P}_\mu, {\cal P}_\nu ] &= 0;  \hspace{4em} [ {\cal M}_{\lambda \mu}, {\cal P}_\nu ] = \delta_{\lambda \nu} {\cal P}_\mu - \delta_{\mu \nu} {\cal P}_\lambda \nonumber\\
\ [{\cal M}_{\lambda \mu}, {\cal M}_{\sigma \tau} ] &= \delta_{\lambda \sigma} {\cal M}_{\mu \tau} - \delta_{\mu \sigma} {\cal M}_{\lambda \tau} - \delta_{\lambda \tau} {\cal M}_{\mu \sigma} + \delta_{\mu \tau} {\cal M}_{\lambda \sigma},
\end{align}
and are symmetries of the action $S$ of \eqref{SGZ} as well,
\beq
[ {\cal P}_\mu, S] = [ {\cal M}_{\lambda \mu}, S] = 0 .
\eeq
More importantly, unlike the algebra of canonical generators of translations and Euclidean rotations,  $\widehat{P}_\mu\ \text{and }  \widehat{M}_{\mu\nu}$, the invariant  Poincar{\'e} generators commute with \emph{all} the charges of phantom symmetries of \eqref{phantomset} defined in~\cite{I}, including the BRST charge $Q_B$, 
\beq
\label{commuteQX}
[ Q_X, {\cal P}_\nu ] = [Q_X, {\cal  M}_{\lambda \mu}  ] = 0 \ \text{for all } Q_X\in\mathfrak{F}\ .
\eeq
Note that \eqref{commutePoincare} and \eqref{commuteQX} imply that the phantom charges, 
\beq\label{tildePoincare}
\widetilde{\cal P}_\mu={\cal P}_\mu-\widehat{\cal P}_\mu\equiv\gamma^{1/2} ( Q_{\bar\varphi,\mu}-Q_{\varphi,\mu})\hspace{5em} \widetilde{\cal M}_{\mu\nu}={\cal M}_{\mu\nu}-\widehat{\cal M}_{\mu\nu}\equiv Q_{M,\mu\nu}\ ,
\eeq
commute with the invariant generators, ${\cal P}_\mu\ \text{and } {\cal M}_{\mu\nu}$, and 
form a closed internal Poincar\'e algebra,
\beqa
\label{hatcommutePoincare}
[ \widetilde{\cal P}_\mu, \widetilde{\cal P}_\nu ] &=& 0;  \ \ \ \ \ \  [ \widetilde{\cal M}_{\lambda \mu}, \widetilde{\cal P}_\nu ] = \delta_{\lambda \nu} \widetilde{\cal P}_\mu - \delta_{\mu \nu} \widetilde{\cal P}_\lambda\nonumber\\
 ~[ \widetilde{\cal M}_{\lambda \mu}, \widetilde{\cal M}_{\sigma \tau} ] &=& \delta_{\lambda \sigma} \widetilde{\cal M}_{\mu \tau} - \delta_{\mu \sigma} \widetilde{\cal M}_{\lambda \tau} - \delta_{\lambda \tau} \widetilde{\cal M}_{\mu \sigma} + \delta_{\mu \tau} \widetilde{\cal M}_{\lambda \sigma} \ .
\eeqa
An indication that these phantom momenta and angular momenta may be unobservable is that they are parts of BRST-doublets. The $\widetilde{M}_{\mu\nu}=Q_{M,\mu\nu}=\{Q_B, Q_{N, \mu\nu}\}$,  given in \eqref{QM}, are exact and $[Q_B,  \widetilde{\cal P}_\mu]= [Q_B, Q_{\bar\varphi,\mu}]=-Q_{\bar\omega,\mu}$ as in \eqref{Qwbar}. Note that the momentum operators of the two Poincar\'e algebras do not simply differ by a BRST-exact charge.

\subsection{Poincar\'e symmetry of physical states}
\label{physPoincare}
In~\cite{I} the physical states of GZ-theory are reconstructed from the physical observables $F\in\mathcal{W}_{\rm phys}$ given by \eqref{W}.  The Jacobi identity and \eqref{commuteQX} imply that $\mathcal{W}_{\rm phys}$ is closed under the invariant Poincar\'e symmetry generated by ${\cal P}_\nu$ and ${\cal  M}_{\mu\nu}$, 
\beq\label{closedPM}
F\in \mathcal{W}_{\rm phys}\ \ \Rightarrow\ \  [{\cal P}_\mu, F]\in\mathcal{W}_{\rm phys}  \ \text{and }   [{\cal M}_{\mu\nu}, F]\in\mathcal{W}_{\rm phys}\ ,
\eeq
since, for instance,  $[Q_X,[{\cal P}_\mu, F]]=[{\cal P}_\mu,[Q_X, F]]+ [F,[{\cal P}_\mu,Q_X]]=0$ for all $F\in \mathcal{W}_{\rm phys}$ and phantom charges $Q_X$.
Further, since invariant and canonical Poincar\'e generators differ only by phantom symmetry charges that commute with $F\in \mathcal{W}_{\rm phys}$, the commutators of either Poincar\'e algebra with physical observables are the same. The set of physical observables  $\mathcal{W}_{\rm phys}$ thus is closed under translations and Euclidean rotations,
 \beq\label{uniqueP}
 [\widehat{\cal P}_\mu, F]=[{\cal P}_\mu, F]\in\mathcal{W}_{\rm phys}  \ \text{and }   [\widehat{\cal M}_{\mu\nu}, F]=[{\cal M}_{\mu\nu}, F]\in\mathcal{W}_{\rm phys}\ ,
 \eeq
and space-time symmetries transform physical observables in $\mathcal{W}_{\rm phys}$ uniquely.

By the reconstruction theorem~\cite{Wightman:1956, I} this translates into a space of physical states with well-defined momentum and angular momentum. Consider for instance a physical state $|F\rangle=F\vac$ corresponding to a physical observable $F\in\mathcal{W}_{\rm phys}$. Using \eqref{symmvac}, an infinitesimal space-time translation changes this state by, 
\beq\label{mapP}
\widehat{\cal P}_\mu|F\rangle=[\widehat{\cal P}_\mu,F]\vac=[{\cal P}_\mu,F]\vac=|[{\cal P}_\mu,F]\rangle\ ,
\eeq
which denotes a \emph{unique} physical state. In this manner translations (and similarly Euclidean rotations) of space-time map the physical subspace of the theory into itself. The commutation relations of \eqref{commutePoincare} imply that these maps form representations of the Poincar\'e algebra. Space-time transformations thus are uniquely realized on the physical Hilbert space of the model. 

\section{The Energy Momentum Tensor}
\label{emGZ}
 It is instructive to investigate the conserved currents associated with space-time symmetries of this model. From the foregoing we expect to find two linearly independent conserved, symmetric and local energy momentum tensors: the Belinfante tensor,  $\widehat{T}_{\mu\nu}(x)$, that couples to a gravitational background and a physical observable, $T_{\mu\nu}(x)$, that is invariant under all phantom symmetries. 

\subsection{The physical energy-momentum tensor $T_{\mu\nu}$}
\label{physT} 
The physical symmetric energy momentum tensor of the theory was derived in~\cite{I} . As in Faddeev-Popov theory it takes the form,
\beq
\label{Tuv}
T_{\mu \nu} = T_{\mu \nu}^{\rm YM} + s \Xi_{\mu \nu}\ ,
\eeq
where  $T_{\mu \nu}^{\rm YM}$ is the (phantom-invariant) energy-momentum tensor of classical Yang-Mills theory,
\beq\label{TYM}
T_{\mu \nu}^{\rm YM}=F_{\mu\kappa}\cdot F_{\nu\kappa}-\frac{\delta_{\mu\nu}}{d} F_{\lambda\kappa}\cdot F_{\lambda\kappa}\ .
\eeq
The additional BRST-exact contribution to $T_{\mu\nu}$ of the GZ-theory is~\cite{I},
\beqa\label{Xi}
\Xi_{\mu \nu} = [ i \p_\mu {\bar c}\cdot A_\nu + \p_\mu \bar\omega_{\kappa} \cdot D_\nu \varphi_{\kappa} - \gamma^{1/2} \text{Tr} D_\mu \bar\omega_\nu ] + [\mu \leftrightarrow \nu]
\nonumber \\
 - \delta_{\mu \nu}(  i \p_\lambda {\bar c}\cdot A_\lambda + \p_\lambda \bar\omega_{\kappa}\cdot D_\lambda \varphi_{\kappa} - \gamma^{1/2} \text{Tr} D_\lambda \bar\omega_\lambda)
\eeqa
with
\beqa\label{sXi}
s\Xi_{\mu \nu}& =&\{Q_B,\Xi_{\mu \nu}\} =  [ i \p_\mu b\cdot A_\nu - i \p_\mu {\bar c}\cdot D_\nu c + \p_\mu \bar\varphi_{\kappa} \cdot D_\nu \varphi_{\kappa} -  \p_\mu \bar \omega_{\kappa}\cdot (D_\nu \omega_{\kappa} + D_\nu c \times \varphi_{\kappa a})
\nonumber \\
&&+ \text{Tr}\gamma^{1/2} (D_\mu \varphi_\nu
 - D_\mu \bar\varphi_\nu - D_\mu c \times \bar\omega_\nu) -\half\delta_{\mu \nu}(2\gamma (N^2-1)+ {\cal L}^{\rm gf})]   + [\mu \leftrightarrow \nu ] \ .
\eeqa
In~\cite{I} we found that $s\Xi_{\mu\nu}$ commutes with all phantom symmetry charges and thus, 
\beq\label{TinW}
T^{\rm YM}_{\mu\nu} \in \mathcal{W}_{\rm phys} \text{ and} \ s\Xi_{\mu\nu} \in \mathcal{W}_{\rm phys} .
\eeq
If the hypothesis of \eqref{hypothesis} formulated in~\cite{I} holds, according to which the expectation value of any BRST-exact observable of $\mathcal{W}_{\rm phys}$ vanishes,  the invariant energy-momentum tensor $T_{\mu\nu}=T^{YM}_{\mu\nu}+s\Xi_{\mu\nu}$ of  GZ-theory is physically equivalent to $T^{YM}_{\mu\nu}$. Although we cannot prove  the hypothesis in general,  the expectation value, $\vev{s \Xi}$ was found~\cite{I} to vanish when the horizon condition of \eqref{horizon} is satisfied,
\be\label{exact0}
\langle s\Xi_{\mu\nu}\rangle=\delta_{\mu\nu}\frac{d-2}{d}\gamma^{1/2}\langle \half\text{Tr} [ D_\kappa (\varphi_\kappa - \bar\varphi_\kappa) - \bar\omega_\kappa\times D_\kappa c  ] -\gamma^{1/2} d (N^2 -1)\rangle =0 \ .
\ee
It is remarkable but physically plausible, that the BRST-exact contribution to the energy-momentum density of the vacuum vanishes only when the horizon condition is satisfied and the effective action is stationary. In~\cite{I} this result was used to relate the trace anomaly of the GZ-theory at the one-loop level to the trace anomaly of Yang-Mills theory. The latter implies $\gamma>0$ and the associated spontaneous breaking of phantom symmetries. 

$s\Xi_{\mu\nu}\in \mathcal{W}_{\rm phys}$ and $s\Psi\in\mathcal{W}_{\rm phys}$, are two examples of  BRST-exact observables whose expectation value was explicitly shown to vanish\cite{I}. If the hypothesis that the expectation value of \emph{any} physical  BRST-exact observable in $\mathcal{W}_{\rm phys}$ vanishes holds true, one recovers the same space of physical observables in GZ-theory as in perturbative Faddeev-Popov theory. GZ-theory in this case is a particular quantization of YM-theory.     

However, since the vacuum state as well as the EM-tensor of this theory are not unique, we need to show that the vacuum of GZ-theory has a well-defined energy-momentum density. Not so surprisingly anymore, this requires the horizon condition to be satisfied.

\subsection{The energy-momentum tensor of Belinfante}
\label{BelinfanteT}
In \eqref{sXi} the auxiliary ghosts  do not transform as vectors in a gravitational background\footnote{See the derivation in~\cite{I} or simply note that for $\gamma=0$ they contribute to $T_{\mu\nu}$ as scalar fields.}. The energy-momentum tensor of Belinfante may be derived by coupling to a gravitational background in a manner that preserves the general covariance of the model also for $\gamma> 0$. The calculation is straightforward but lengthy and has been relegated to Appendix~\ref{Belinfante}. The conserved and symmetric EM-tensor $\widehat{T}_{\mu\nu}$ of Belinfante for the GZ-theory one obtains is,
\beq\label{hatTuv}
\widehat{T}_{\mu\nu}=T_{\mu\nu} +\Delta_{\mu\nu}
\eeq
where  $\Delta_{\mu\nu}(x)$ is itself a symmetric and conserved current,
\beqa\label{Delta}
\Delta_{\mu\nu}&=&\Big[\gamma (N^2-1)\delta_{\mu\nu} + \p_\kappa(H_{\kappa\mu\nu}+H_{\kappa\nu\mu}-H_{\mu\nu\kappa})+\\
&&\hspace{3em}+\gamma^{1/2}{\rm Tr}\Big(\half\delta_{\mu\nu} \p_\kappa (\varphi_\kappa-\bar\varphi_\kappa)-\p_\nu\varphi_\mu+D_\mu\bar\varphi_\nu +\bar\omega_\mu \times D_\nu c \Big)\Big]  + [\mu\leftrightarrow\nu] \ ,\nonumber
\eeqa
with 
\beq\label{Htensor}
H_{\mu\nu\kappa}=\half[\bar\varphi_{\mu} \cdot D_\nu\varphi_{\kappa}+\bar\omega_{\mu}\cdot(\varphi_{\kappa}\times D_\nu c)-\bar\omega_{\mu} \cdot D_\nu\omega_{\kappa}-\p_\nu\bar\varphi_{\mu} \cdot\varphi_{\kappa}+\p_\nu\bar\omega_\mu\cdot \omega_{\kappa} ]
\eeq
The contributions $H_{\mu\nu\kappa}$ arise because the auxiliary ghosts of the model canonically transform as vectors under Euclidean rotations. This difference of \eqref{Delta} to the physical EM-tensor therefore does not even vanish for $\gamma=0$.

We show in Appendix~\ref{Belinfante} that the divergence $\p_\nu \Delta_{\mu\nu}$ is proportional to equations of motion of the  unphysical fields of the theory and thus vanishes on-shell. The canonical energy momentum tensor of Belinfante as well as the physical one are thus conserved on-shell. However, the Belinfante energy-momentum tensor does not commute with a number of phantom symmetry charges, and in particular does not commute with the BRST charge $Q_B$. It thus is not a physical observable of the theory. $\langle \widehat{T}_{\mu\nu}\rangle $ nevertheless, by definition, is the response to a change in the (classical) gravitational background. 

The vacuum energy-momentum  density in fact  \emph{is} well defined, since
\beq\label{fakevacen}
\langle \Delta_{\mu\nu}\rangle=\frac{2\delta_{\mu\nu}}{d} [\gamma^{1/2}{\rm Tr}\langle D_\kappa\bar\varphi_\kappa +\bar\omega_\kappa \times D_\kappa c \rangle +\gamma d (N^2-1)]=\frac{2\delta_{\mu\nu}}{d}\gamma^{1/2}\langle s \text{Tr} D_\mu \bar\omega_\mu \rangle=0\ ,
\eeq
when \eqref{halfhorizon} is satisfied. If the horizon condition holds, we thus have that,
 \beq\label{equalTs}
\langle\widehat T_{\mu\nu}\rangle=\vev{T_{\mu\nu}}=\vev{T_{\mu\nu}^{\rm YM}}\ ,
\eeq
and the vacuum energy-momentum density is well defined in this theory.

\section{Ward identities for $T_{\mu \nu}$ and $M_{\lambda \mu \nu}$} 
\label{WardId}
 
The symmetric physical energy-momentum tensor $T_{\mu \nu}$, given by Eqs.~(\ref{Tuv},\ \ref{TYM}\ ,\ref{Xi}), provides non-trivial Ward identities for gauge-invariant and therefore \emph{physical} correlation functions. In this section we derive and explore these basic physical Ward identities of the theory.  

 The divergence of $T_{\mu\nu}$ depends linearly on the equation of motion $\delta S \over \delta \Phi_i$.  Rearranging terms one obtains the identity
 \beq
 \label{Tidentity}
 \p_\mu T_{\mu \nu}(x) = - \mathfrak{g}_\nu (x) S,
 \eeq
 where $\mathfrak{g}_\nu(x)$ is the differential operator
 \drop{\beqa
 
 \mathfrak{g}_\nu(x) & \equiv  & \p_\nu A_\mu {\delta \over \delta A_\mu}  - \p_\mu \left( A_\nu {\delta \over \delta A_\mu}  \right)  + \p_\nu b  {\delta \over \delta b} + \p_\nu c {\delta \over \delta c} + \p_\nu \bar c  {\delta \over \delta \bar c}
 \nonumber \\
&&  +  {\delta \over \delta \phi_\mu} \p_\nu \phi_\mu + \p_\nu \bar\phi_\mu  {\delta \over \delta \bar\phi_\mu}  + \p_\nu \omega_\mu {\delta \over \delta \omega_\mu} + \p_\nu \bar\omega_\mu  {\delta  \over \delta \bar\omega_\mu}.
 \eeqa}
\beqa\label{gsubnu}
\mathfrak{g}_\nu(x)  & = & \p_\nu A_\mu \cdot {\delta \over \delta A_\mu}  - \p_\mu \left( A_\nu \cdot {\delta \over \delta A_\mu}  \right) + \p_\nu b\cdot {\delta  \over \delta b}  + \p_\nu c\cdot {\delta  \over \delta c} + \p_\nu \bar c \cdot {\delta  \over \delta \bar c}
 \nonumber \\
 &&   + \p_\nu \omega_\mu \cdot{\delta  \over \delta \omega_\mu} + \p_\nu \bar\omega_\mu  \cdot{\delta  \over \delta \bar\omega_\mu}  + \p_\nu \varphi_\mu\cdot {\delta  \over \delta \varphi_\mu}+ \p_\nu \bar\varphi_\mu\cdot{\delta  \over \delta \bar\varphi_\mu}
 \nonumber \\
&&   +  \gamma^{1/2}\Tr  \left({\delta  \over \delta \bar\varphi_\nu} - {\delta  \over \delta \varphi_\nu}-i \bar\varphi_\nu\times{\delta  \over \delta b}-i\bar\omega_\nu\times{\delta  \over \delta \bar c}  \right).
 \eeqa

 The first two terms of $\mathfrak{g}_\nu(x)$ may be rewritten,
 \beq
 \label{gsubAS}
 \p_\nu A_\mu \cdot {\delta \over \delta A_\mu}  - \p_\mu \left( A_\nu\cdot {\delta \over \delta A_\mu}  \right) = F_{\nu \mu} \cdot {\delta \over \delta A_\mu}  - A_\nu\cdot D_\mu  {\delta \over \delta A_\mu}.
 \eeq
Note that for $\gamma=0$, the terms of \eqref{gsubnu} (except for the total divergence  $\p_\mu \left( A_\nu {\delta S \over \delta A_\mu} \right)$) all generate a space-time translation of the fields.
 
To arrive at a Ward identity, we insert the classical relation of \eqref{Tidentity} into a matrix element,
 \beq
 \vev{{\cal O} \ \p_\mu T_{\mu \nu}(x) } = - \vev{{\cal O} \ \mathfrak{g}_\nu (x) S }= \int d\Phi \ {\cal O} \ \mathfrak{g}_\nu(x) \exp(-S)\ ,
 \eeq
 where $\cal O$ is a generic functional of the fields.  Functional integration by parts then gives the Ward identity,
\beq
\label{momWI}
 \vev{  {\cal O} \ \p_\mu T_{\mu \nu}(x) } = - \vev{\mathfrak{g}_\nu (x) {\cal O}}\ .
\eeq

Because the energy-momentum tensor is symmetric, $T_{\mu \nu} = T_{\nu \mu}$, the angular momentum current,
\beq
M_{\lambda \mu \nu} \equiv x_\lambda T_{\mu \nu} - x_\mu T_{\lambda \nu},
\eeq
satisfies
\beq
\p_\nu M_{\lambda \mu \nu} =  x_\lambda \p_\nu T_{\mu \nu} - x_\mu \p_\nu T_{\lambda \nu}.
\eeq
By the same reasoning as before, this gives the corresponding Ward identity
\beq
\label{angmomWI}
 \vev{ {\cal O} \ \p_\nu M_{\lambda \mu \nu}(x)  } = - \vev{ (x_\lambda \mathfrak{g}_\mu (x) - x_\mu \mathfrak{g}_\lambda (x)) {\cal O} }.
\eeq
 
 To obtain an identity among physical quantities only, consider \eqref{momWI} for an observable ${\cal O}=I(A)$, where $I(A)$ is a gauge-invariant functional of the gauge-connection $A$ only, $I({^g}\!A) = I(A)$.  \eqref{momWI} for this special case simplifies to,
 \beq
 \vev{ I(A)\ \p_\mu T_{\mu \nu} } = \vev{F_{\mu\nu}\cdot  {\delta I[A] \over \delta A_\mu}},
 \eeq
 since the contribution from the second term of \eqref{gsubAS} vanishes, because $I(A)$ is gauge invariant and $D_\mu {\delta I(A) \over \delta A_\mu} = 0$. Decomposing the invariant energy-momentum tensor  into the classical one of Yang-Mills and the BRST-exact contribution $\Xi_{\mu\nu}$ of \eqref{Tuv},  one finally has, 
 \beq\label{physWard}
 \vev{ I(A)\ \p_\mu T_{\mu \nu}^{\rm YM}(x) } + \vev{ s ( I(A)\ \p_\mu \Xi_{\mu \nu})} = \vev{  F_{\mu \nu}(x) \cdot {\delta I(A) \over \delta A_\mu}(x) } .
 \eeq
Note that \eqref{physWard} uses that $\Xi_{\mu\nu}$ anti-commutes with the phantom symmetry $Q_{\bar\omega,\mu}$ and that $s\Xi_{\mu\nu}\in \mathcal{W}_{\rm phys}$ is an invariant operator and thus,
\begin{align}\label{proofcommute}
\p_\mu s \Xi_{\mu \nu}&=[\widehat{\cal P}_\mu, \{Q_B, \Xi_{\mu \nu}\} ] \nonumber\\&=\{Q_B,[\widehat{\cal P}_\mu, \Xi_{\mu \nu}]\} +\{[\widehat{\cal P}_\mu,Q_B], \Xi_{\mu \nu}\}\nonumber\\
&=s\p_\mu \Xi_{\mu \nu}-\gamma^{1/2}\{Q_{\bar\omega,\mu},\Xi_{\mu \nu}\} \nonumber\\
&=s\p_\mu \Xi_{\mu \nu}\ .
\end{align}
The first line of \eqref{proofcommute} expresses $\p_\mu s \Xi_{\mu \nu}$ in terms of functional derivative operators. The second is Jacobi's identity and \eqref{commPQ} then gives the third line. The final result in \eqref{proofcommute} uses that $\Xi_{\mu\nu}$, given by \eqref{Xi}, anti-commutes with the phantom charge $Q_{\bar\omega,\mu}$.
Since $\mathcal{W}_{\rm phys}$ is closed under translations and $s\Xi_{\mu \nu}\in\mathcal{W}_{\rm phys}$, we have that $\p_\mu s \Xi_{\mu \nu}= s \p_\mu\Xi_{\mu \nu}\in \mathcal{W}_{\rm phys}$.

According to the hypothesis of \eqref{hypothesis}, the expectation value of the BRST-exact physical observable in \eqref{physWard} vanishes (just as in perturbative Faddeev-Popov theory). We conclude that gauge invariant physical operators $I(A)=I({^g}\!A)$ satisfy the Ward identity,
\beq\label{physWard1}
 \vev{ I(A)\ \p_\mu T_{\mu \nu}^{\rm YM}(x)  } = \vev{  F_{\mu \nu}(x) \cdot {\delta I(A) \over \delta A_\mu}(x) } ,
 \eeq
that involves gauge-invariant observables only. 
Let us now use an entirely different approach to derive the same physical Ward identity. 

\subsection{An alternate direct point of view}  
\label{alternateview}
         
A simple calculation of the divergence of the gauge-invariant Yang-Mills part of the energy-momentum tensor of \eqref{TYM} yields,
\beq\label{YMdiv}
\p_\mu T_{\mu \nu}^{\rm YM} = F_{\mu \nu}\cdot {\delta S^{\rm YM} \over \delta A_\mu} .
\eeq
Decomposing the total action of \eqref{SGZ} in $S = S^{\rm YM} + S^{\rm gf}$ one may rewrite \eqref{YMdiv} as,
\beq\label{YMdiv1}
\p_\mu T_{\mu \nu}^{\rm YM} + F_{\mu \nu}\cdot {\delta S^{\rm gf} \over \delta A_\mu}  =F_{\mu \nu}\cdot {\delta S \over \delta A_\mu}.
\eeq
Because $S^{\rm gf}$ is  BRST-exact and because $Q_B$ commutes with  $F_{\mu\nu}\cdot\frac{\delta}{\delta A_\mu}$, the second term of \eqref{YMdiv1} is a BRST-exact expression,
\beq
\label{sexactterminWI}
 F_{\mu \nu}\cdot{\delta S^{\rm gf} \over \delta A_\mu} = s \Sigma_\nu,
\eeq
where
\begin{align}\label{Sigma}
\Sigma_\nu & \equiv F_{\mu \nu}\cdot  (i \p_\mu \bar c +\varphi_\lambda\times \p_\mu \bar\omega_\lambda)+\gamma^{1/2} \Tr  \bar\omega_\mu\times F_{\mu \nu}
\nonumber \\
s \Sigma_\nu & = F_{\mu \nu}\cdot [ i \p_\mu b + \varphi_\lambda\times \p_\mu \bar\varphi_\lambda  + \omega_\lambda\times\p_\mu \bar\omega_\lambda+(i \p_\mu \bar c + \varphi_\lambda\times\p_\mu \bar\omega_\lambda)\times c]  
\nonumber  \\
&\hspace{3em} + \gamma^{1/2} \Tr \left[(\bar\varphi_\mu -  \varphi_\mu)\times F_{\mu\nu}- \bar\omega_\mu\times (c \times F_{\mu \nu})\right].
\end{align}
We here used that $s F_{\mu \nu}= - c \times F_{\mu \nu}$.  This gives,
\beq
\p_\mu T_{\mu \nu}^{\rm YM} + s \Sigma_\nu = F_{\mu \nu}\cdot{\delta S \over \delta A_\mu} .
\eeq

Inserting this identity into a matrix element with the generic functional ${\cal O}$ and  performing the partial functional integration,
\beq\label{funcint}
\vev{{\cal O}\ F_{\mu \nu}\cdot{\delta S \over \delta A_\mu(x)}}=\vev{ F_{\mu \nu}\cdot{\delta {\cal O}\over \delta A_\mu(x)}}\ ,
\eeq
one obtains an alternate form of the Ward identity\footnote{Note that \eqref{altWard} is derived without calculating or even identifying the complete energy-momentum tensor $T_{\mu \nu}$.},
 \beq\label{altWard}
 \vev{{\cal O}\ \p_\mu T_{\mu \nu}^{\rm YM}(x) } + \vev{ {\cal O}\  s \Sigma_\nu(x) } = \vev{\ F_{\mu \nu}\cdot{\delta {\cal O} \over \delta A_\mu(x)}}\ .
 \eeq
For \eqref{momWI} and \eqref{altWard} to be valid for \emph{any} (not necessarily  physical) functional ${\cal O}$, we must have that,
\beq\label{equivWard}
s\Sigma_\nu(x)=s \p_\mu\Xi_{\mu\nu}+ \mathfrak{g}_\nu(x) S +F_{\mu\nu}\cdot{\delta S \over \delta A_\mu(x)}\ .
\eeq
Explicitly verifying \eqref{equivWard} is straightforward, if tedious.
It is more interesting to consider \eqref{altWard} for a gauge-invariant physical observable ${\cal O}=I(A)=I({^g}\!A)$. The Ward identity in this case takes the form of \eqref{physWard1} except for the BRST-exact term,
 \beq\label{physWard2}
 \vev{I(A)\ \p_\mu T_{\mu \nu}^{\rm YM}(x)} + \vev{ s (I(A)\ \Sigma_\nu)} = \vev{  F_{\mu \nu}(x) \cdot {\delta I(A) \over \delta A_\mu}(x) } .
 \eeq
To be consistent with \eqref{physWard1}, the expectation value of the BRST-exact functional in \eqref{physWard2} must vanish.  This is compatible with the hypothesis of \eqref{hypothesis}, because $s\Sigma_\nu$ is itself a (BRST-exact) physical observable that commutes with \emph{all} phantom symmetries,\footnote{To see this note that in \eqref{sexactterminWI}, the gauge-fixing part of the action $S^{\rm gf}\in\mathcal{W}_{\rm phys}$. The only phantom charges that do not commute with $A_\nu$ and $\frac{\delta}{\delta A_\nu}$ are the generators of rigid gauge transformations $Q_C$  given in \eqref{QC} and the nilpotent BRST-charge $Q_B$ of \eqref{QB}. Both evidently leave  $s\Sigma_\nu$ invariant.}    $s\Sigma_\nu\in \mathcal{W}_{\rm phys}$. 

\section{Discussion and Conclusions}
\label{diss}

In this article we demonstrated that energy, momentum and angular momentum are well-defined symmetry generators on the space $\mathcal{W}_{\rm phys}$ of physical observables defined in \eqref{W} and on the associated physical Hilbert space.  We related two linearly independent Poincar\'e algebras of conserved charges: the set of generators of space-time symmetries $\{\widehat{\cal P}_\mu,\widehat{\cal M}_{\mu\nu}\}$ given by Eqs.~(\ref{hatP}--\ref{sdensity}) that do not commute with the BRST-charge $Q_B$ and other phantom charges, and the set of invariant generators that do, $\{{\cal P}_\mu,{\cal M}_{\mu\nu}\}$ of~\cite{I}.  The two Poincar\'e algebras differ by the phantom charges given in \eqref{P} and \eqref{M} and therefore transform any physical observable $F\in\mathcal{W}_{\rm phys}$ in the same way. We showed that the space of physical observables $\mathcal{W}_{\rm phys}$ is closed under isometries of Euclidean space-time and that these symmetries are uniquely generated on $\mathcal{W}_{\rm phys}$.   

The Hilbert space of the theory is reconstructed from the physical observables  of $\mathcal{W}_{\rm phys}$.  With a non-vanishing Gribov parameter $\gamma\neq 0$, the ground state of the theory breaks the BRST-symmetry spontaneously but remains translationally and rotationally invariant.   Space-time transformations act uniquely on the reconstructed invariant Hilbert space as in \eqref{mapP}. The vacuum breaks a number of phantom symmetries spontaneously, but has vanishing ghost number and is colorless.  The $Q_{R,\mu a}$ of \eqref{QR} generate interesting unbroken phantom symmetries, that ensure (see \eqref{auxFP}) the equality of auxiliary fermi-ghost and Faddeev-Popov (FP) ghost propagators for any background connection.        

Corresponding to its two Poincar\'e algebras, the theory has two linearly independent conserved symmetric EM-tensors: that of Belinfante, $\widehat{T}_{\mu\nu}$ of \eqref{EMGZ}, which couples to a gravitational background but does not commute with all phantom charges (also not BRST) and another, $T_{\mu\nu}$,  invariant under all phantom symmetries (including BRST). The two differ by the symmetric tensor $\Delta_{\mu\nu}$ given in \eqref{Delta}. The latter is conserved up to equations of motion of unphysical fields. 

It was found, \eqref{fakevacen},  that $\vev{\Delta_{\mu\nu}}$ is proportional to the expectation value of a BRST-exact density that vanishes~\cite{I}, courtesy of the  horizon condition, provided the symmetry generated by $Q_{R,\mu a}$ is unbroken.  The vacuum energy-momentum  density of the GZ-theory in this sense  is well-defined even though its BRST-symmetry is spontaneously broken.   

In support of the hypothesis of \eqref{hypothesis}, the criterion of \eqref{criterion} was used in~\cite{I} to prove that the BRST-exact parts of the invariant energy-momentum tensor $T_{\mu\nu}$ and of the action $S$  have vanishing expectation value.  Although we cannot prove the hypothesis \eqref{hypothesis} at present, we have found no evidence to the contrary.  The GZ-theory thus is a candidate for a consistent quantization of classical YM-theory despite its spontaneously broken BRST symmetry.  The hypothesis in  particular implies that gauge-invariant observables satisfy the same physical Ward identity with the Yang-Mills energy-momentum tensor as in perturbative Faddeev-Popov theory.

The hypothesis of \eqref{hypothesis} may be correct but too restrictive.  Indeed there are examples of BRST-exact operators that are not in $\mathcal{W}_{\rm phys}$, whose expectation value was explicitly shown to vanish~\cite{I}. Although not essential for the consistency of the theory, a less restrictive algebraic characterization of BRST-exact operators whose expectation value vanishes is desireable.  However, we have not been able to algebraically distinguish between the operator $s\text{Tr} D_\mu \bar \omega_\nu(x)$, whose expectation value vanishes, and $s\text{Tr} \partial_\mu \bar \omega_\nu(x)$ with  non-vanishing expectation value.

In summary, the Gribov mass parameter $\gamma>0$ and associated mass gap of the GZ-theory arise from the spontaneous dynamical breakdown of global symmetries as in other models without an explicit low-energy (mass) scale. The corresponding gap equation is the horizon condition of \eqref{horizon}. The  spontaneously broken symmetries in this theory are unphysical in the sense that all observables are invariants. The breaking of these phantom symmetries differs from a conventional Higgs mechanism in that some of the broken symmetries (including BRST) are fermionic.  Despite the broken BRST symmetry, we find  that this construction is surprisingly consistent.  We have a symmetric energy momentum tensor $T_{\mu\nu}$ that is a physical observable, and the corresponding charges generate space-time isometries on the subspace $\mathcal{W}_{\rm phys}$ of physical observables and on the associated reconstructed physical Hilbert space.  Poincar\'e symmetry is recovered in the physical sector of this theory,  even though the canonical generators of translation of \eqref{hatP} do not commute with the BRST-charge $Q_B$.

We speculate that the non-perturbative quantization of a non-abelian gauge theory maintains translational invariance \emph{and} BRST symmetry in the physical sector only. A non-trivial commutation relation between the BRST-charge and the momentum, like \eqref{commPQ} in the GZ-theory,  implies that the ground state either is not translationally invariant or breaks BRST (see \eqref{choice}). The GZ-quantization breaks BRST symmetry spontaneously but maintains translational and rotational invariance.  Lattice regularization apparently explores the complimentary scenario of strictly maintaining gauge symmetry at the expense of manifest Poincar\'e symmetry. On the lattice there is no global BRST-symmetry that respects lattice symmetries~\cite{Neuberger:1987} as well as the compactness of the structure group~\cite{Schaden:1998,Smekal:2007}.  Constraining configurations of a lattice gauge theory to the fundamental modular region by, for instance,  selecting a maximal tree of specified link variables, even breaks the discrete translation and hypercubic symmetries.  This also holds for any  gauge of non-vanishing degree~\cite{Sharpe:1984}. Gauge-invariant correlators  of course do not depend on the selected maximal tree and the explicitly broken Poincar\'e symmetry of the lattice gauge theory presumably is restored in the continuum limit for gauge invariant observables.   The regularized GZ-theory on the other hand maintains Poincar\'e symmetry but breaks BRST spontaneously. However we found support for the hypothesis  of~\eqref{hypothesis} that BRST is regained where it counts~\cite{I}, and that expectation values of BRST-exact physical observables of $\mathcal{W}_{\rm phys}$ indeed vanish in the continuum theory.

{\vspace{2em}\noindent\bf Acknowlegements: }MS would like to thank members of the physics department of New York University for their hospitality. 
 
\appendix
\section{Some Phantom Symmetries and their Generators}
\label{phantomsymm}
Phantom symmetries by definition~\cite{I} are unobservable continuous symmetries  Any observable of the theory commutes with all phantom charges. 
We denote charges and generators of phantom symmetries by $Q_X$, with the subscript $X$ identifying the symmetry. For reference we here collect the generators of phantom symmetries used in this article. We refer the interested reader to~\cite{I} for the definitions of all the phantom symmetry generators in the set,
\beq\label{phantoms}
\mathfrak{F}=\{Q_B,Q_\mathcal{N},Q_+, Q_{\bar\varphi,\mu},Q_{\bar\omega,\mu},Q^a_G,Q^a_C,Q^a_{NL},Q^a_{\bar c},Q_{R,\mu a},Q_{S,\mu a},Q_{E,\mu\nu\,ab},Q_{F,\mu\nu\,ab},Q^a_{U,\mu b},Q^a_{V,\mu b}\}\  .
\eeq

The most  important of the phantom symmetries is BRST. It is a nilpotent symmetry generated by the fermionic charge $Q_B$ given in \eqref{QB}. Nilpotency means that the BRST charge anticommutes with itself,
\beq\label{nilpotency}
\{Q_B,Q_B\}=0
\eeq
as can be verified using \eqref{QB}. All physical observables are expected to be BRST-invariant. Although it governs the structure and renormalizability of a gauge theory, the BRST symmetry is  fundamentally unobservable and thus the most prominent phantom (symmetry). Together with the ghost number, $Q_{\cal N}$, $Q_B$  forms a BRST-doublet. It turns out that all phantom symmetries are BRST-doublets.  

Contrary to Faddeev-Popov theory in Landau gauge, GZ-theory does not appear\footnote{We at least could not find the corresponding generators.} to possess a nilpotent anti-BRST charge that (anti-)commutes with $Q_B$ nor an SL(2,R) symmetry~\cite{I}. 

The Poincar{\'e} algebra of phantom charges of GZ-theory arises due to invariance of \eqref{Lgf} with respect to internal shifts and rotations of  auxiliary fields. The variation $\delta\bar\varphi^a_{\mu b}=\bar\epsilon_\mu \delta^a_b$ with constant $\bar\epsilon_\mu=0$ is one of them. This shift is generated by,
\beq\label{Qphibar}
Q_{\bar\varphi,\mu} \equiv \int d^dx \ \text{Tr} {\delta \over \delta \bar\varphi_\mu} .
\eeq

The commutator of $Q_{\bar\varphi, \mu}$  with the BRST charge $Q_B$ of \eqref{QB}  gives the associated phantom charge,  
\beq
\label{Qwbar}
Q_{\bar\omega,\mu}=-s Q_{\bar\varphi,\mu}=-[Q_B,Q_{\bar\varphi,\mu}] \equiv \int d^dx \ \text{Tr}{\delta \over \delta \bar\omega_\mu} ,
\eeq
and $(Q_{\bar\varphi,\mu},Q_{\bar\omega,\mu})$ are a BRST-doublet of color-singlet charges. Note that invariance of physical operators under $Q_B$ \emph{and} $Q_{\bar\omega,\mu}$ implies their invariance under $Q_B+ a_\mu Q_{\bar\omega,\mu}$ for an arbitrary constant vector $a_\mu$. The choice of origin in the definition of the BRST-charge $Q_B$ in \eqref{QB} thus is of no import for operators that commute with the phantom symmetry generator $Q_{\bar\omega,\mu}$.

For transverse configurations,  $\p_\mu A_\mu=0$, the GZ-action of \eqref{SGZ} by inspection  also is  invariant with respect to the shift   $\delta \omega^a_{\mu b}=\epsilon_\mu \delta^a_b$ . This color singlet phantom symmetry is generated by the charge,
\beq\label{Qw}
Q_{\omega, \mu} \equiv \int d^dx \ {\mathfrak{q}}_\mu,
\eeq
with the density,
\beq\label{tq}
\mathfrak{q}_\mu \equiv  \text{Tr}\Big[ {\delta \over \delta {\omega}_\mu}  + i {\bar\omega}_\mu\times{\delta \over \delta b}\Big]\ . 
\eeq
The BRST-variation of $Q_{\omega, \mu}$ is,
\beq\label{Qphi}
Q_{\varphi, \mu} =s Q_{\omega, \mu}=\{Q_B,Q_{\omega, \mu}\}\equiv \int d^dx\ \mathfrak{p}_\mu 
\eeq
with the density,
\beq
\label{tp}
\mathfrak{p}_\mu = s\mathfrak{q}_\mu= \text{Tr}\Big[{\delta  \over \delta \varphi_\mu} + i  \bar\varphi_\mu\times   {\delta  \over \delta b}  + i {\bar\omega}_\mu \times  {\delta  \over \delta {\bar c}}\Big]\ .
\eeq
$(Q_{\omega, \mu},Q_{\varphi, \mu})$ thus is another BRST-doublet of color-singlet phantom generators.

Finally, at $\gamma=0$,  ${\cal L}^{\rm gf}$ evidently is invariant under the internal symmetry $\delta \omega^a_{\mu b}=\epsilon_{\mu\nu} \varphi^a_{\nu b}; \delta \bar\varphi^a_{\mu b}=\epsilon_{\mu\nu} \bar\omega^a_{\nu b}$ with antisymmetric  $\epsilon_{\mu\nu}=-\epsilon_{\nu\mu}$ that rotates anti-commuting and commuting auxiliary fields into each other. This phantom (super-)symmetry persists for  $\gamma> 0$ in an extended form generated by, 
\beq
\label{QN}
Q_{N, \mu\nu}\equiv  
\int d^dx \ \left[ \varphi_{\nu}\cdot {\delta  \over \delta \omega_{\mu}}
+ {\bar\omega}_{\nu} \cdot{\delta  \over \delta \bar\varphi_{\mu}} 
 + \gamma^{1/2} x_\mu \mathfrak{q}_\nu \right] - [\mu \leftrightarrow \nu].
\eeq
The BRST-variation of $Q_{N, \mu\nu}$ in \eqref{QN} gives the generators of an internal O(4) symmetry of the auxiliary ghost, 
\beq\label{QM}
Q_{M, \mu \nu} =\{Q_B, Q_{N, \mu\nu}\} \equiv  \int d^dx \ \left[  \mathfrak{s}_{\mu \nu} +\gamma^{1/2} x_\mu (\mathfrak{p}_\nu-\text{Tr} {\delta  \over \delta \bar\varphi_\nu})\right] -[\mu\leftrightarrow\nu]
\eeq
with densities  $\mathfrak{s}_{\mu \nu}$ and $\mathfrak{p}_\mu$ given in \eqref{sdensity} and \eqref{tp} respectively.

In the GZ-theory, rigid color rotations are unbroken and generated by~\cite{Schaden:1994,Mader:2014,I},
\beqa
\label{QC}
Q_C=& \equiv & \int d^dx \Big[ A_\mu \times  {\delta \over \delta A_\mu} + c \times  {\delta \over \delta c} + \bar c \times  {\delta \over \delta \bar c} + b \times  {\delta \over \delta b}
+ \varphi_{\mu a} \times  {\delta \over \delta \varphi_{\mu a}} + \bar\varphi_{\mu a} \times  {\delta \over \delta \bar\varphi_{\mu a}}\nonumber\\
&&\hspace{-2em} + \omega_{\mu a} \times  {\delta \over \delta \omega_{\mu a}} + \bar\omega_{\mu a} \times  {\delta \over \delta \bar\omega_{\mu a}}
 + \varphi^a_\mu \widetilde\times  {\delta \over \delta \varphi^a_\mu} + \bar\varphi^a_\mu \widetilde\times  {\delta \over \delta \bar\varphi^a_\mu} + \omega^a_\mu \widetilde\times  {\delta \over \delta \omega^a_\mu} + \bar\omega^a_\mu \widetilde\times  {\delta \over \delta \bar\omega^a_\mu} \Big]\ .
\eeqa
 
Faddeev-Popov theory in Landau gauge and GZ-theory share the remarkable property~\cite{Blasi:1991}, that the color charge $Q_C$ is BRST-exact. With the BRST-charge of \eqref{QB} we have that~\cite{Schaden:1994}, 
\beq
\label{QCexact}
Q_C^a =   \{Q_B, Q_G^a\} \ ,
\eeq
where
\beqa\label{QG}
\hspace{-2em}Q_G \equiv \int d^dx \ \left[  -  {\delta \over \delta c} +  {\bar c} \times  {\delta \over \delta  b}  +  \varphi_{\mu a}  \times  {\delta \over \delta \omega_{\mu a}} +  {\bar\omega}_{\mu a}\times {\delta \over \delta \bar\varphi_{\mu a}}   + \varphi^a_\mu \widetilde\times {\delta \over \delta \omega^a_\mu}  + {\bar\omega}^a_\mu \widetilde\times  {\delta \over \delta \bar\varphi^a_\mu}  \right]\ .
\eeqa
The pairs $(Q_G^a,Q_C^a)$  thus are BRST doublets. 

The GZ-theory in addition possesses a set of phantom symmetries in the adjoint that mix auxiliary vector ghosts with  FP-ghosts. Inspection of ${\cal L}^{\rm gf}$ in \eqref{Lgf} reveals the invariance, $\delta \bar c^a=i\eps_{\mu b} \bar\omega^a_{\mu b}, \delta\omega^a_{\mu b}=\eps_{\mu b} c^a$. This phantom symmetry is generated by the charges, 
\beq\label{QR}
Q_{R,\mu a} =\int d^dx [c\cdot {\delta  \over \delta \omega_{\mu a}}
+i\bar\omega_{\mu a}\cdot{\delta  \over \delta \bar c}]\ .
\eeq
of vanishing ghost number. We believe these to be unbroken phantom symmetries.  Commuting with the BRST-generator $Q_B$ reveals another set of phantom generators in the adjoint representation,
\beq\label{QS}
Q_{S,\mu a} =-[Q_B,Q_{R,\mu a}]= \int d^dx [\half (c\times c)\cdot {\delta  \over \delta \omega_{\mu a}}+c\cdot {\delta  \over \delta \varphi_{\mu a}}
-i\bar\varphi_{\mu a}\cdot {\delta  \over \delta \bar c}-i\gamma^{1/2} x_\mu {\delta  \over \delta \bar c^a}]\ ,
\eeq
which are readily verified to commute with the GZ-action of \eqref{SGZ}.

For the definition of the other phantom charges of \eqref{phantoms} and a more complete discussion we refer the interested reader to Appendix~A of~\cite{I}\ .

 \section{The canonical EM-Tensor of the GZ-model}
\label{Belinfante}
We here compute the canonical EM-Tensor $\widehat{T}_{\mu \nu}$  of Belinfante from the action $S$ of \eqref{SGZ} written in general coordinates. 
To simplify some manipulations and for bookkeeping purposes we introduce contra- and co-variantly transforming vector fields  and a manifestly scalar (graded) BRST-transformation $s_0 F\equiv[Q_0,F]_\pm$  by,
\beq\label{Q0}
Q_0=\left. Q_B\right|_{\gamma=0} = \int d^dx \sqrt{-g}\ \left[ D_\mu c \cdot {\delta \over \delta A_\mu} - \half (c \times c)\cdot {\delta \over \delta c} + b\cdot {\delta \over \delta {\bar c}} + \omega_{\mu}\cdot {\delta \over \delta \varphi_{\mu}} + \bar\varphi^\mu\cdot {\delta \over \delta {\bar\omega}^\mu} \right].
\eeq
Although nilpotent, $Q_0$ does not generate a symmetry\footnote{We merely use $Q_0$ to facilitate calculation of the canonical EM-tensor.  In this article we consider only exact symmetries of the action and are not otherwise concerned with this operator, which generates an explicitly (albeit softly) broken symmetry~\cite{Sorella:2004,Dudal:2005,Capri:2014}.}  of the GZ-action for  $\gamma>0$, but contrary to $Q_B$ of \eqref{QB}, has the advantage of not  depending explicitly on coordinates.  In terms of $s_0$, the gauge-fixing part of the GZ-action of \eqref{Lgf}  in general coordinates may be written,
\beq\label{Lgf1}
\L^{\rm gf}=\gamma^{1/2}\sqrt{-g}\text{ Tr} (g^{\mu\nu}D_\nu\varphi_\mu)-\gamma\sqrt{-g} (N^2-1)+s_0\Psi_0\ ,
 \eeq
with
\be\label{a1}
\Psi_0=\sqrt{-g}\{ ig^{\mu\nu}(\p_\mu {\bar c})\cdot A_\nu +\bar\omega^{\mu;\nu}\cdot  D_\nu \varphi_{\mu}-\gamma^{1/2}\text{ Tr} (A_\mu\times\bar\omega^\mu )\}\ .
\ee
Note that $\bar\varphi^{\mu a}_b=s_0\bar \omega^{\mu a}_b$ are covariantly transforming auxiliary vector ghosts, whereas $\omega^a_{\mu b}=s_0\varphi^a_{\mu b}$ transform contravariantly. These assignments are compatible with $Q_0$ in \eqref{Q0} under coordinate transformations [$s_0 A_\mu=D_\mu c$ is similarly compatible].  However, $s_0$ is not a symmetry of the action and a part of $\mathcal{L}^{\rm gf}$ is not $s_0$-exact.   

The covariant derivatives \footnote{Following convention~\cite{MTW}, covariant derivatives in this Appendix are denoted by a semicolon and partial derivatives by a comma, that is $X_{\mu;\kappa}\equiv\nabla_\kappa X_{\mu},\ X_{\mu,\kappa}\equiv\p_\kappa X_\mu$.} of \eqref{a1} are defined as,
\begin{subequations}
\label{a2}
\begin{align}
\bar\omega^{\mu a;\nu}_b&\equiv g^{\nu\rho}\bar\omega^{\mu a}_{b\,,\rho}+g^{\nu\sigma}\Gamma^{\mu}_{\rho\sigma}\bar\omega^{ \rho a}_b\label{a2a}\\
(D_\nu \varphi_{\mu b})^a&\equiv \varphi^a_{\mu b;\nu}+(A_\nu\times \varphi_{\mu b})^a= \varphi^a_{\mu b,\nu}+(A_\nu\times \varphi_{\mu b})^a-\Gamma^{\rho}_{\mu\nu}\varphi^a_{\rho b}\label{a2b}
\end{align}
\end{subequations}
with a Christoffel symbol,
\be\label{a3}
\Gamma^\rho_{\mu\nu}=\Gamma^\rho_{\nu\mu}\equiv\half g^{\rho\sigma}\left(g_{\sigma\mu,\nu}+g_{\sigma\nu,\mu}-g_{\mu\nu,\sigma}\right),
\ee
that is symmetric in its lower indices. [$g_{\mu\nu;\rho}=g_{\mu\nu,\rho}-g_{\nu\sigma} \Gamma^\sigma_{\mu\rho}-g_{\mu\sigma} \Gamma^\sigma_{\nu\rho}=0$.] 
  
Let us first obtain the contribution $s_0\widehat{\Xi}_{\mu\nu}$ to the energy-momentum tensor arising from the term $s_0 \Psi_0$ in $\L^{\rm gf}$ of \eqref{Lgf1}, where $\widehat{\Xi}_{\mu\nu}$ is defined in \eqref{a9}.
We wish to compute $\Psi_0$ for,
\beq\label{expand}
g^{\mu\nu}(x)=\eta^{\mu\nu}+2\epsilon^{\mu\nu}(x),
\eeq 
to leading order\footnote{$A \approx B$ here means that  expressions $A$ and $B$ coincide to leading non-trivial order in $\epsilon^{\mu\nu}$. By taking the leading deviation from the Minkowski metric to be $2\epsilon_{\mu\nu}$ we avoid a proliferation of factors $1/2$.}  in the deviation $\epsilon^{\mu\nu}$ from the metric $\eta_{\mu\nu}$ . To this order we have,
\begin{subequations}
\label{a5}
\begin{align}
\label{gmunue}
g_{\mu\nu}&\approx \eta_{\mu\nu}-2\epsilon_{\mu\nu}\\
\label{detge}
\sqrt{-g}&\approx 1-\epsilon^{\mu\nu}\eta_{\mu\nu}\\
\label{Christoffele}
\Gamma^\rho_{\mu\nu}&\approx \epsilon_{\mu\nu}^{\ \ ,\rho}-\epsilon^\rho_{\ \mu,\nu}-\epsilon^\rho_{\ \nu,\mu}\ .
\end{align}
\end{subequations}
To leading order,  indices in terms proportional to $\epsilon$ have here been  "pulled" by the space-time independent Minkowski metric $\eta_{\mu\nu}$. Inserting the expressions of \eqref{a5} in \eqref{a2} one obtains that to leading order in $\epsilon^{\mu\nu}$,
\begin{subequations}
\label{a6}
\begin{align}
\bar\omega^{\mu a;\nu}_b&\approx\bar\omega^{\mu a,\nu}_b+2\epsilon^{\nu\rho}\omega^{\mu a}_{b\,,\rho} -\left(\epsilon^{\mu\nu,\rho}+\epsilon^{\rho\mu,\nu}-\epsilon^{\rho\nu,\mu}\right) \bar\omega^a_{\rho b}\label{a6a}\\
(D_\nu \varphi_{\mu b})^a&\approx\varphi^a_{\mu b,\nu}+(A_\nu\times \varphi_{\mu b})^a+\left(\epsilon_{\rho\mu,\nu}+\epsilon_{\rho\nu,\mu}-\epsilon_{\mu\nu,\rho}\right)\varphi^{\rho a}_b\label{a6b}
\end{align}
\end{subequations}
With these expressions and those of \eqref{a5}, the terms proportional to $\epsilon^{\mu\nu}$ of $\Psi_0$ are,
\begin{align}
\label{a9}
\Psi_0&\approx\left.\Psi_0\right|_{g_{\mu\nu}=\eta_{\mu\nu}}+\epsilon^{\mu\nu}\widehat{\Xi}_{\mu\nu}\ \ \text{with, }\nonumber\\
\widehat{\Xi}_{\mu\nu}&=i  \bar c_{,\mu} \cdot A_\nu+i  \bar c_{,\nu}\cdot  A_\mu-\eta_{\mu\nu} i  \bar c^{,\rho}\cdot A_\rho+(G_{\rho\nu\mu}+G_{\rho\mu\nu}-G_{\mu\nu\rho})^{,\rho}\nonumber\\
&+\bar\omega^\rho_{\ ,\nu}\cdot D_\mu\varphi_{\rho}+ \bar\omega^\rho_{\ ,\mu} \cdot D_\nu\varphi_{\rho} -\eta_{\mu\nu}\left(\bar\omega^{\rho ,\sigma} \cdot D_\sigma \varphi_{\rho}-\gamma^{1/2} \text{ Tr} (A_\rho\times\bar\omega^\rho)\right)\ ,
\end{align}
where,
\be\label{a10}
G_{\mu\nu\rho}=G_{\nu\mu\rho}=\half(\bar\omega_{\mu}\cdot D_\nu \varphi_{\rho}-\bar\omega_{\mu\,,\nu}\cdot \varphi_{\rho}+\bar\omega_{\nu} \cdot D_\mu \varphi_{\rho}-\bar\omega_{\nu\,,\mu}\cdot\varphi_{\rho})\ ,
\ee
arises from the Christoffel symbols of \eqref{Christoffele} associated with the vector auxiliary ghosts. With the covariant derivative of \eqref{a6b}, the contribution to the canonical energy momentum tensor from the remaining terms of $\L^{\rm gf}$ in \eqref{Lgf1} is,
\be\label{newEM}
\eta_{\mu\nu} \gamma d (N^2-1)+\gamma^{1/2} \text{ Tr} \left[ A_\mu\times\varphi_\nu+A_\nu\times\varphi_\mu-\eta_{\mu\nu} (A_\rho\times \varphi^\rho)\right]\ .
\ee
Note that the total (covariant) divergence in \eqref{Lgf1} does not contribute to the energy-momentum tensor. Adding a term $-\gamma^{1/2} \sqrt{-g}\, \text{Tr} \bar\omega^\mu_{\ ;\mu}$ to $\Psi_0$ in \eqref{a1} for an explicitly (gauge-)covariant divergence does not alter the canonical energy-momentum tensor either. 

The canonical symmetric energy momentum tensor of the GZ-model thus is,
\be\label{EMGZ}
\widehat{T}_{\mu\nu}=T^{YM}_{\mu\nu}+\eta_{\mu\nu} \gamma d (N^2-1)+\gamma^{1/2} \text{ Tr} \left[ A_\mu\times\varphi_\nu+A_\nu\times\varphi_\mu-\eta_{\mu\nu} (A_\rho\times \varphi^\rho)\right]+ s_0\widehat{\Xi}_{\mu\nu}\ ,
\ee
with $\widehat{\Xi}_{\mu\nu}$ given in \eqref{a9} and $s_0$ generated by $Q_0$ of \eqref{Q0}.

The canonical energy momentum tensor of \eqref{EMGZ} differs from the invariant energy momentum tensor $T_{\mu\nu}$ obtained in~\cite{I} and  reproduced in \eqref{Tuv}. To compute the difference efficiently, we rewrite the BRST-exact term of $T_{\mu\nu}$ as an $s_0$ variation and a remainder,
\beq\label{Tuv1}
T_{\mu \nu} =T^{YM}_{\mu\nu} +\delta_{\mu \nu}\gamma(d-2)(N^2-1) +\gamma^{1/2}\text{Tr}\left[D_\nu \varphi_\mu+D_\mu \varphi_\nu - \delta_{\mu \nu}D_\kappa \varphi_\kappa\right]+s_0 \Xi_{\mu \nu}\ .
\eeq
  
The difference, $\Delta_{\mu\nu}$ of \eqref{hatTuv}, between the canonical- and the invariant-  symmetric energy momentum tensors of the GZ-theory thus is,
\be\label{diff}
\Delta_{\mu\nu}=\widehat{T}_{\mu\nu}-T_{\mu\nu}=s_0(\widehat{\Xi}_{\mu\nu}- \Xi_{\mu\nu})+2\delta_{\mu\nu}\gamma(N^2-1)-\gamma^{1/2}\text{ Tr}\left[ \varphi_{\mu,\nu}+\varphi_{\nu,\mu} - \eta_{\mu \nu} \varphi_\rho^{\ ,\rho}\right]\ .
\ee
Written out explicitly, the first contribution is,
\beqa\label{Xidiff}
s_0(\widehat{\Xi}_{\mu\nu}-\Xi_{\mu \nu}) &=&s_0(G_{\rho\nu\mu}+G_{\rho\mu\nu}-G_{\mu\nu\rho})^{,\rho}+\gamma^{1/2} s_0\text{ Tr}\left[  D_\mu \bar\omega_\nu+ D_\nu\bar\omega_\mu-\eta_{\mu\nu} \bar\omega^\rho_{\ ,\rho}\right]\\
&&\hspace{-6em}=\left[(H_{\rho\nu\mu}+H_{\rho\mu\nu}-H_{\mu\nu\rho})^{,\rho}+ \gamma^{1/2}\text{ Tr}(D_\mu \bar\varphi_\nu+(D_\mu c)\times\bar\omega_\nu-\half\eta_{\mu\nu}\bar\varphi^{\rho}_{\ ,\rho}) \right]+[\mu\leftrightarrow\nu]\ ,\nonumber
\eeqa
with $H_{\mu\nu\rho}$ defined in \eqref{Htensor} and satisfying,
\beq\label{Hdef}
s_0 G_{\mu\nu\rho}=H_{\mu\nu\rho}+H_{\nu\mu\rho}\ .
\eeq

If the canonical and invariant EM-tensors are both conserved, the divergence of their difference should be proportional to equations of motion of unphysical fields only. It is interesting to verify this explicitly. The definition of \eqref{a10} implies that, 
\begin{align}
\label{a11}
(G_{\rho\nu\mu}+G_{\rho\mu\nu}-G_{\mu\nu\rho})^{,\rho\nu}&=G_{\rho\nu\mu}^{\ \ \ ,\rho\nu}=(\bar\omega_\rho \cdot D_\nu \varphi_{\mu}-\bar\omega_{\rho,\nu}\cdot \varphi_{\mu})^{,\rho\nu}\nonumber\\
&=(\bar\omega_\rho \cdot \partial^\nu (D_\nu  \varphi_{\mu})+\bar\omega_\rho^{,\nu} \cdot  (A_\nu\times\varphi_{\mu})-\bar\omega_{\rho,\nu}^{\ ,\nu}\cdot \varphi_{\mu})^{,\rho}\nonumber\\
&=(\bar\omega_\nu \cdot \partial^\rho (D_\rho  \varphi_{\mu})-(D_\rho\partial^\rho\bar\omega_{\nu})\cdot \varphi_{\mu})^{,\nu}\ , 
\end{align}
and the divergence of $\Delta_{\mu\nu}$ can therefore be written,
\begin{align}\label{diffdiv}
\Delta_{\mu\nu}^{\ \ ,\nu}&=s_0 (\widehat{\Xi}_{\mu\nu}-{\Xi}_{\mu\nu})^{,\nu}-\gamma^{1/2}\text{Tr}\ \partial_\nu \varphi_\mu^{\ ,\nu}\nonumber\\
&=\text{ Tr}\p^\nu (s_0 [\bar\omega_\nu\cdot \partial^\rho D_\rho  \varphi_\mu- (D_\rho\partial^\rho\bar\omega_\nu)\cdot \varphi_\mu+\gamma^{1/2}(A_\mu\times \bar\omega_\nu+ D_\nu\bar\omega_\mu)]-\gamma^{1/2}\p_\nu\varphi_\mu) \nonumber\\
&=\partial_\nu (\frac{\delta S}{\delta\varphi_\nu}\cdot \varphi_\mu-\frac{\delta S}{\delta\omega_\nu}\cdot\omega_\mu-\bar\varphi^\nu\cdot\frac{\delta S}{\delta\bar\varphi^\mu}-\bar\omega^\nu\cdot \frac{\delta S}{\delta\bar\omega^\mu})\nonumber\\
&\hspace{5em}+\gamma^{1/2}\text{Tr}\,(i\frac{\delta S}{\delta b}\times\bar\varphi_\mu -i\frac{\delta S}{\delta\bar c}\times\bar\omega_\mu-\frac{\delta S}{\delta\varphi^\mu}+\frac{\delta S}{\delta\bar\varphi^\mu})\nonumber\\
&=\partial_\nu (\varphi_\mu\cdot \frac{\delta S}{\delta\varphi_\nu}+\omega_\mu\cdot \frac{\delta S}{\delta\omega_\nu}-\bar\varphi^\nu\cdot \frac{\delta S}{\delta\bar\varphi^\mu}-\bar\omega^\nu\cdot \frac{\delta S}{\delta\bar\omega^\mu})+ \gamma^{1/2}(\text{Tr}\frac{\delta S}{\delta\bar\varphi^\mu}-\mathfrak{p}_\mu S)\ .
\end{align}
Here we made use of the equations of motion of $\omega,\bar\varphi,\bar c$ and the Nakanishi Lautrup field $b$,
\be\label{eom}
\frac{\delta S}{\delta\omega_\mu}=-D_\nu\partial^\nu\bar\omega^{\mu}\ ,\quad \frac{\delta S}{\delta\bar\varphi^\mu}=-\partial^\nu D_\nu  \varphi_{\mu}+\gamma^{1/2}  A_\mu\times\ , \quad \frac{\delta S}{\delta b}=-i\partial^\nu A_\nu \ , \quad \frac{\delta S}{\delta \bar c}=i\partial^\nu D_\nu c\ ,
\ee
as well as those of $\varphi$ and $\bar\omega$ written in the form,
\begin{subequations}
\label{eoms}
\begin{align}
\frac{\delta S}{\delta\varphi_\mu} &=-s_0 D_\nu\partial^\nu\bar\omega^\mu+\gamma^{1/2} A^\mu\times &=&-D_\nu c\times \bar\omega^{\mu,\nu}- D_\nu \partial^\nu\bar\varphi^\mu+\gamma^{1/2} A^\mu\times
\label{eomphi}\\
\frac{\delta S}{\delta\bar\omega^\mu}&=s_0\partial^\nu D_\nu\varphi_\mu-\gamma^{1/2} D_\mu c\times &=&(D_\nu c\times\varphi_\mu)^{,\nu} +\partial^\nu D_\nu \bar\omega_\mu- \gamma^{1/2} D_\mu c\times\label{eomomegab}\ .
\end{align}
\end{subequations}
\eqref{diffdiv} shows that the divergence of the difference $\Delta_{\mu\nu}=\widehat{T}_{\mu\nu}-T_{\mu\nu}$ of the canonical- and invariant- energy momentum tensors is proportional to equations of motion of unphysical fields. Since we proved that the invariant EM-tensor, $T_{\mu\nu}$, is conserved in~\cite{I}, both energy momentum tensors are conserved onshell. \eqref{diffdiv} also shows that $\Delta_{\mu\nu}$ is the conserved Noether current corresponding to the phantom symmetry generated by $\widetilde{\cal P}_\mu \equiv Q_{\bar\varphi,\mu}-Q_{\varphi,\mu}$. Since $\Delta_{\mu\nu}$ is  a symmetric conserved tensor, the currents $x_\mu \Delta_{\rho\nu}-x_\nu \Delta_{\rho\mu}$ are also conserved and correspond to the Noether currents of the  
symmetries generated by the phantom charges $\widetilde{\cal  M}_{\mu\nu}\equiv Q_{M,\mu\nu}$ of \eqref{QM}.

\end{document}